\documentclass[aps,prb,twocolumn]{revtex4}

\usepackage{graphicx}
\usepackage{amsmath}

\begin{document}

%\title{Electron-phonon energy exchange in the presence of the superconducting proximity effect}
\title{Phase sensitive electron-phonon coupling in a superconducting proximity structure}

\author{T.~T. Heikkil\"a}
\email[]{Tero.Heikkila@tkk.fi} \affiliation{Low Temperature
Laboratory, Helsinki University of Technology, P.O. Box 5100
FIN-02015 TKK, Finland}

\author{Francesco Giazotto}

\affiliation{NEST CNR-INFM and Scuola Normale Superiore, I-56126
Pisa, Italy}

\newcommand{\tempnote}[1]%
   {\begingroup{\it (NOTE: #1)}\endgroup}
\newcommand{\sgn}{{\rm sgn}}

\date{\today}

\begin{abstract}
We study the role of the superconducting proximity effect on the
electron-phonon energy exchange in diffusive normal metals (N)
attached to superconductors (S). The proximity effect modifies the
spectral response of the normal metal, in particular the local
density of states. This leads to a weakening of the electron-phonon
energy relaxation. We show that the effect is easily observable with
modern thermometry methods, and predict that it can be tuned in
structures connected to multiple superconductors by adjusting the
phase difference between superconducting order parameters at the two
NS interfaces.
\end{abstract}


\maketitle

\section{Introduction}

The state of an electron system subject to driving depends on three
ingredients: on the external driving, on the response functions of
the system, and on internal and external relaxation within the
system and out of the system. For sufficiently strong but constant
driving, the system can be brought out of equilibrium from its
surroundings, in which case the steady state can be determined from
a balance between the driving and the relaxation. By far, the most
relevant relaxation mechanism for electrons in metals is caused by
the coupling between electrons and phonons. If the dominant
relaxation mechanism is known to a high accuracy, the result of such
heat balance can be used to study the driving. An example of such a
procedure takes place in hot-electron thermal radiation
detectors,\cite{giazotto06,giazotto08} which rely on the changes in
the electron temperature due to changes in the amount of radiation
coupling to the device and driving the electron system. Typically
such devices require accurate thermometry through an easily
measurable observable that is sensitive to the electron temperature.
One such frequently employed thermometer is based on a junction
between normal metals and superconductors.

\begin{figure}[h]
\centering
\includegraphics[width=\columnwidth,clip]{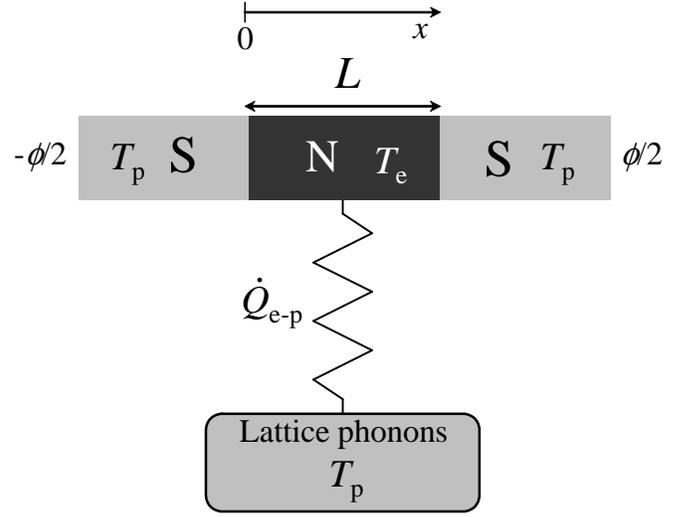}
\caption{Schematic structure considered in this paper: the
superconducting electrodes (assumed to be large and residing at the
phonon temperature $T_p$) induce a proximity effect into the normal
metal. This proximity effect changes the spectrum of the electronic
excitations and thereby also the electron-phonon energy exchange
$\dot Q_{e-p}$. The spectrum can be controlled via the phase
difference $\phi$ of the superconducting order parameter. This
allows one to explore the effect as shown below in
Fig.~\ref{fig:heatbalance}.} \label{fig:scheme}
\end{figure}

When a normal metal is brought in contact with superconductors, the
superconducting order parameter leaks out to the normal side. This
superconducting proximity effect \cite{degennes63} changes the
spectral response of the system \cite{virtanen07} but it also
affects the relaxation mechanisms. In this paper we study how this
effect manifests itself in the electron-phonon coupling and consider
the schematic structure shown in Fig.~\ref{fig:scheme}. Some of the
features induced by the proximity effect are the changes in the
local density of states \cite{joyez08} and a finite pair amplitude
inside the normal metal. For example, the density of states obtains
a minigap whose size $E_g(\phi)$ depends on the phase difference
$\phi$ between the superconductors:\cite{zhou99} at $\phi=0$ for
$E_{\rm Th} \ll \Delta$, $E_g \approx 3.12 E_{\rm Th}$ whereas
$E_g=0$ at $\phi=\pi$. Here $E_{\rm Th}=\hbar D/L^2$ is the Thouless
energy of the normal metal piece with a diffusion constant $D$ and
length $L$ and $\Delta$ is the energy gap of the superconductor.

By utilizing quasiclassical equations, we show how the collision
integrals describing the electron-phonon coupling are changed by the
proximity effect. We consider in particular the resulting effects on
the electron-phonon relaxation rate, and the heat power flowing
between the two systems. Our examples are calculated for a SNS
system where the normal metal is sandwiched between two
superconductors. In this case the proximity-induced effects can also
be tuned by adjusting the phase difference between the order
parameters of the two superconducting contacts.

This paper is organized as follows. At first we derive the general
diffusive-limit electron-phonon collision integrals from Usadel
equation \cite{usadel70} for Keldysh Green's functions.\cite{kopnin}
Then we study especially the case when the electron system can be
described with Fermi function at a local temperature $T_e$ and the
phonon system with Bose function at temperature $T_{p}$. For small
deviations $|T_e-T_{p}| \ll T_e$ the collisions can be described
with a relaxation rate, whose dependence on energy, temperature and
$\phi$ are shown for an example geometry. Then we concentrate on the
energy current between the two systems, the quantity entering heat
balance equations. This quantity is well defined for arbitrary
temperature differences between the electron and phonon systems. Its
dependence on position, length of the normal metal and $\phi$ are
exemplified for example geometries. Finally we suggest how the phase
dependence can be measured in a system easily manufactured with
present experimental techniques.

\section{Theoretical framework}

When describing nonequilibrium superconductivity, it is convenient
to concentrate separately on two types of
excitations:\cite{schmidschoen} "longitudinal", described by the
antisymmetric part $f_L=f(-E)-f(E)$ of the electron energy
distribution function with respect to the chemical potential of the
superconductor, and "transverse", described by the symmetric part
$f_T=1-f(E)-f(-E)$. The previous describes "thermal" excitations
whereas the latter is relevant for charged excitations. The detailed
forms of the collision integrals for the two types of excitations
are derived in the Appendix. At quasiequilibrium described by an
effective temperature $T_e$, the most relevant type of excitations
are the longitudinal ones. In this case, assuming the elastic mean
free path exceeds the phonon wavelength,\cite{giazotto06} the
collision integral is given by
\begin{equation}\label{eq:quasieqcollil}
\begin{split}
I_L= \lambda \int d\omega \omega^2 \sgn(\omega)
K(\epsilon,\omega)R(\epsilon,\omega;T_e,T_p).
\end{split}
\end{equation}
Here $\lambda$ is the electron-phonon coupling constant, the kernel
$$K(\epsilon,\omega)=N(\epsilon)N(\epsilon+\omega)-F_c(\epsilon)F_c(\epsilon+\omega)-F_s(\epsilon)F_s(\epsilon+\omega)$$
describes the changes in the spectrum of the junction, $N(\epsilon)$
is the (local) reduced density of states, and $F_c(\epsilon)$ and
$F_s(\epsilon)$ are projections of the pair amplitude (anomalous
function) as described in Eq.~\eqref{eq:ffuncs}. The part depending
on the phonon and electron temperatures is in quasiequilibrium
\begin{equation*}
\begin{split}
&R(\epsilon,\omega;T_e,T_p)=\tanh\left(\frac{\epsilon}{2 k_B
T_e}\right)\tanh\left(\frac{\epsilon+\omega}{2 k_B
T_e}\right)+\\&\coth\left(\frac{\omega}{2 k_B T_{p}}\right)
\left[\tanh\left(\frac{\epsilon+\omega}{2k_B
T_e}\right)-\tanh\left(\frac{\epsilon}{2 k_B T_e}\right)\right]-1.
\end{split}
\end{equation*}
The kernel $K(\epsilon,\omega)$ is in general position dependent,
and thereby it makes also the collision integral depend on the
position. Once this collision integral is known, it can be inserted
in a kinetic equation, such as those presented in
Ref.~\onlinecite{virtanen07}.

In what follows, we describe the electron-phonon scattering inside
the SNS junction in terms of the scattering rate and the heat
current flowing between the electron and phonon systems. In order to
find the density of states and the anomalous function inside the
normal region of the SNS junction, we solve numerically the Usadel
equation \cite{usadel70,virtanen07,hammer07,numnote}
\begin{equation}\label{eq:usadel}
D\nabla \cdot \left(\hat{g}^R \nabla \hat{g}^R\right) =
[-i(\epsilon+ i \gamma) \hat{\tau}_3,\hat{g}^R],
\end{equation}
where $\gamma$ is a small positive parameter describing inelastic
scattering and the retarded Green's function $\hat g^R$ satisfies
the normalization condition $(\hat{g}^R)^2=\hat{1}$. We assume that
the NS interfaces are clean and that $\hat{g}^R$ obtains the form of
bulk Green's function $\hat{g}^R_{\rm
bulk}=\epsilon_+/\sqrt{\epsilon_+^2-\Delta^2} \hat{\tau}_3 +
\Delta/\sqrt{\epsilon_+^2-\Delta^2}[\cos(\phi)
i\hat{\tau}_2+\sin(\phi) i \hat \tau_1]$ at these boundaries. Here
$\epsilon_+=\epsilon+i\gamma$, $\Delta$ is the absolute value of the
superconducting order parameter, $\phi$ is its phase and $\hat
\tau_i$ are the Pauli matrices in Nambu space. For
lower-transparency junctions the proximity effect and thereby the
effects described below will be reduced.\cite{hammer07}

\section{Scattering rate}

Close to equilibrium, the electron-phonon scattering can be
described with an energy dependent scattering rate.\cite{kaplan76}
The latter can be obtained by assuming that the Keldysh part of the
scattering self-energy is related to the retarded (R) and advanced
(A) parts via $\hat{\Sigma}^K = (\hat{\Sigma}^R-\hat{\Sigma}^A)f_L$
and similarly for the Green's function, $\hat{g}^K \approx
(\hat{g}^R-\hat{g}^A)(f_L+\delta f_L)$. This gives us $I_L \approx
\Gamma_{\rm e-p}^{\rm SNS} \delta f_L$ with
%\begin{widetext}
\begin{equation}
\begin{split}
&\Gamma_{\rm e-p}^{\rm SNS} = \frac{1}{2} {\rm
Tr}[(\hat{g}^R-\hat{g}^A)(\hat{\Sigma}^R-\hat{\Sigma}^A)]\\& =
\lambda \int d\omega \omega^2 \sgn(\omega) K(\epsilon,\omega)
\frac{\cosh\left(\frac{\epsilon}{2 k_B
T}\right)}{\sinh\left(\frac{\omega}{2 k_B T}\right)
\cosh\left(\frac{\epsilon+\omega}{2 k_B T}\right)}.
\end{split}
\end{equation}
%\end{widetext}
In the absence of superconductivity, $N(\epsilon) \approx 1$ is
approximately a constant and $F_c=F_s \equiv 0$. In this case at
$\epsilon=0$ we get $\Gamma_{\rm e-p}^N=7 \lambda \zeta(3) (k_B
T)^3$.\cite{rammer98}

\begin{figure}[h]
\centering
\includegraphics[width=\columnwidth,clip]{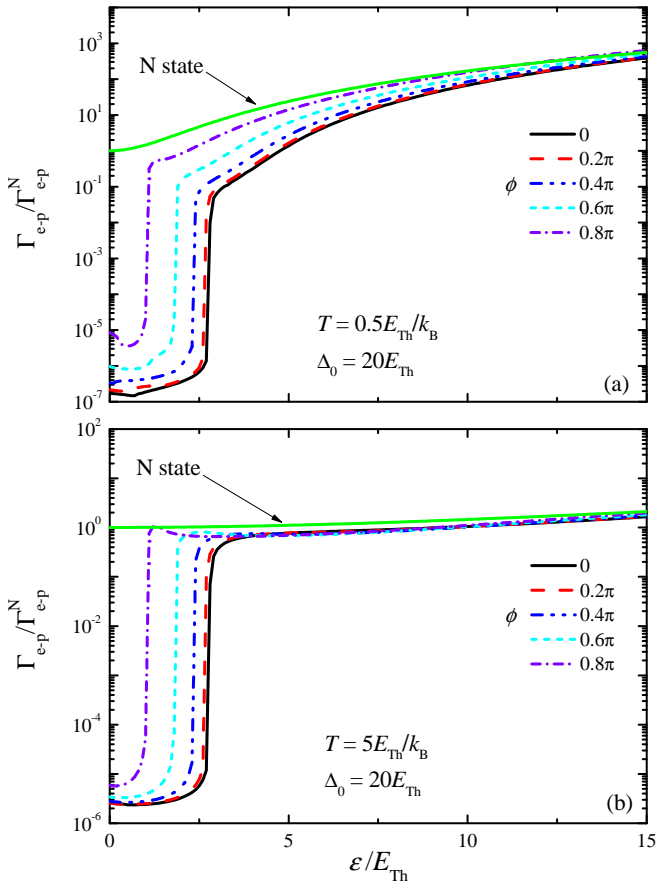}
\caption{(color online) Phase and energy dependence of the
electron-phonon scattering rate $\Gamma_{\rm e-p}$ normalized to the
normal-state zero-energy result $\Gamma_{\rm e-p}^{\rm N}=7\lambda
\zeta(3)(k_B T)^3$ at two different temperatures (top: $k_B T=E_{\rm
Th}/2$ and bottom: $k_B T=5 E_{\rm Th}$). We have assumed
$\Delta_0=20 E_{\rm Th}$ and averaged the rates across the
normal-metal piece. The solid line shows the energy dependence
obtained in the normal case for which $K(\epsilon,\omega)=1$. Note
the logarithmic scale of the figure. The exact low-energy values in
the presence of the proximity effect depend on the magnitude of the
inelastic scattering parameter $\gamma$ in Eq.~\eqref{eq:usadel}.
Here and throughout we have chosen $\gamma=10^{-5}E_{\rm Th}$.}
\label{fig:rate}
\end{figure}

The scattering rate is plotted in Fig.~\ref{fig:rate} as a function
of energy for a few phases $\phi$ and at two different temperatures.
The plotted rate is averaged across the weak link,
\begin{equation*}
\Gamma_{\rm e-p} = \frac{1}{L}\int_0^L dx \Gamma_{\rm e-p}^{\rm
SNS}(x),
\end{equation*}
and normalized to $\Gamma_{\rm e-p}^N$ to illustrate the corrections
due to the proximity effect. At energies $\epsilon > k_B T$ the rate
rises as $\epsilon^3$ in the absence of superconductivity. The main
modification due to the proximity effect is the phase-dependent
minigap, which causes a huge drop in the scattering rates. Moreover,
above the minigap the rates are slightly lower than in the absence
of the proximity effect
--- this is due to the fact that within a thermal coherence length
$\xi_T = \sqrt{\hbar D/(2\pi k_B T)}$ from the NS interfaces
electron-phonon coupling is weakened at any value of the phase (see
Fig.~\ref{fig:posdep}).

\section{Heat current}

The most relevant quantity in describing the electron-phonon
interaction is the heat current associated with a temperature
difference in the electron and phonon systems. This quantity can be
described also far from equilibrium, i.e., for arbitrary difference
between the electron and phonon temperatures. The heat current
density is obtained by multiplying $I_L$ with energy $\epsilon$ and
the normal-state density of states $\nu_F$ at Fermi energy, and
integrating over the energy:
%\begin{widetext}
\begin{align}
P_{\rm \rm e-p}=\lambda \nu_F \int d \epsilon d\omega \epsilon
\omega^2 \sgn(\omega)K(\epsilon,\omega) R(\epsilon,\omega;T_e,T_p).
\end{align}
%\end{widetext}
In the normal state, the integral can be evaluated analytically,
with the result\cite{wellstood94}
\begin{equation}
P_{\rm \rm e-p}^{N}=\Sigma (T_p^5-T_e^5),
\end{equation}
where $\Sigma=24\zeta(5) \nu_F k_B^5\lambda$. Measured values of
$\Sigma$ in different metals are tabulated in
Ref.~\onlinecite{giazotto06}.

In most cases, the heat diffusion within the normal metal is much
stronger than out from it. In this case the variations in the
temperature of the normal metal become small and the average
temperature is determined by the rate integrated over the volume of
the junction, i.e.,
\begin{equation}
\dot{Q}_{\rm e-p} = A\int_0^L dx P_{\rm \rm e-p}(x).
\end{equation}
Here $A$ is the cross section of the normal metal. In the normal
case $P_{\rm e-p}^N$ is independent of position and thus
$\dot{Q}_{\rm e-p}^N=P_{\rm e-p}^N \Omega$, where $\Omega=AL$ is the
volume of the normal metal.

\begin{figure}[h]
\centering
\includegraphics[width=\columnwidth,clip]{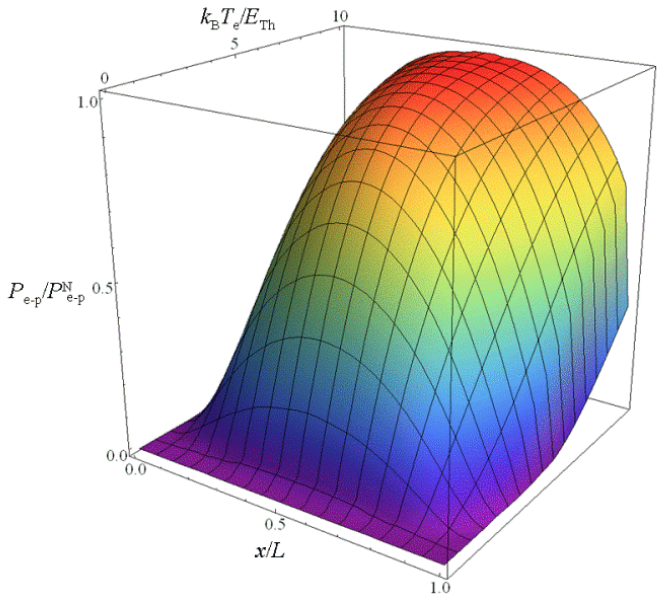}
\caption{(color online) Position and temperature dependent heat
current density $P_{\rm e-p}$ between the electron and phonon
systems inside the normal part of the SNS system of length $L$. Here
$\phi=0$, $\Delta=20E_{\rm Th}$, $T_{p}=0$ and the heat current
density has been normalized to the normal-state value $P_{\rm
e-p}^N=\Sigma T_e^5$.} \label{fig:posdep}
\end{figure}

Proximity effect changes the electron-phonon heat current
drastically. Most of the changes are limited within a thermal
coherence length $\xi_T=\sqrt{\hbar D/(2 \pi k_B T)}$ from the
superconductors. This is illustrated in Fig.~\ref{fig:posdep}, which
shows $P_{\rm \rm e-p}$ normalized to $P_{\rm \rm e-p}^N$ as a
function of position and temperature. Note that changing $T_e$
inside the normal metal does not affect the superconducting gap as
the superconductors are assumed to be at the phonon temperature.

\begin{figure}[t]
\centering
\includegraphics[width=\columnwidth,clip]{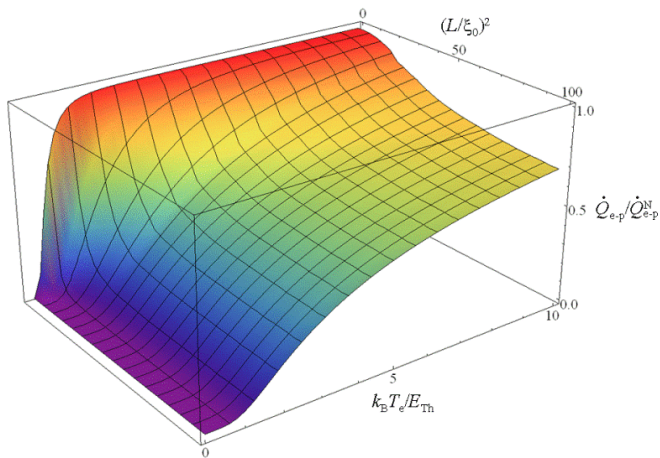}
\caption{(color online) Dependence of the heat current on electron
temperature and the length $L$ of the junction compared to the
zero-temperature coherence length $\xi_0=\sqrt{\hbar D/\Delta}$.
Here $\phi=0$ and $T_{p}=0$. } \label{fig:lengthdep}
\end{figure}

A relevant factor for the SNS system is the relation of the energy
gap inside the superconductor to the Thouless energy of the weak
link. This ratio equals to the square of the ratio between the
length of the junction and the superconducting coherence length.
Figure \ref{fig:lengthdep} shows the position-averaged heat current
as a function of electron temperature for different lengths of the
junction. At low temperatures below the minigap, $k_B T \lesssim
E_g$, the heat current is vanishingly small because of the lack of
states inside the normal metal. However, also at higher temperatures
there is some reduction, whose measurement is well within reach of
state-of-the-art thermometry.

\begin{figure}[h]
\centering
\includegraphics[width=\columnwidth,clip]{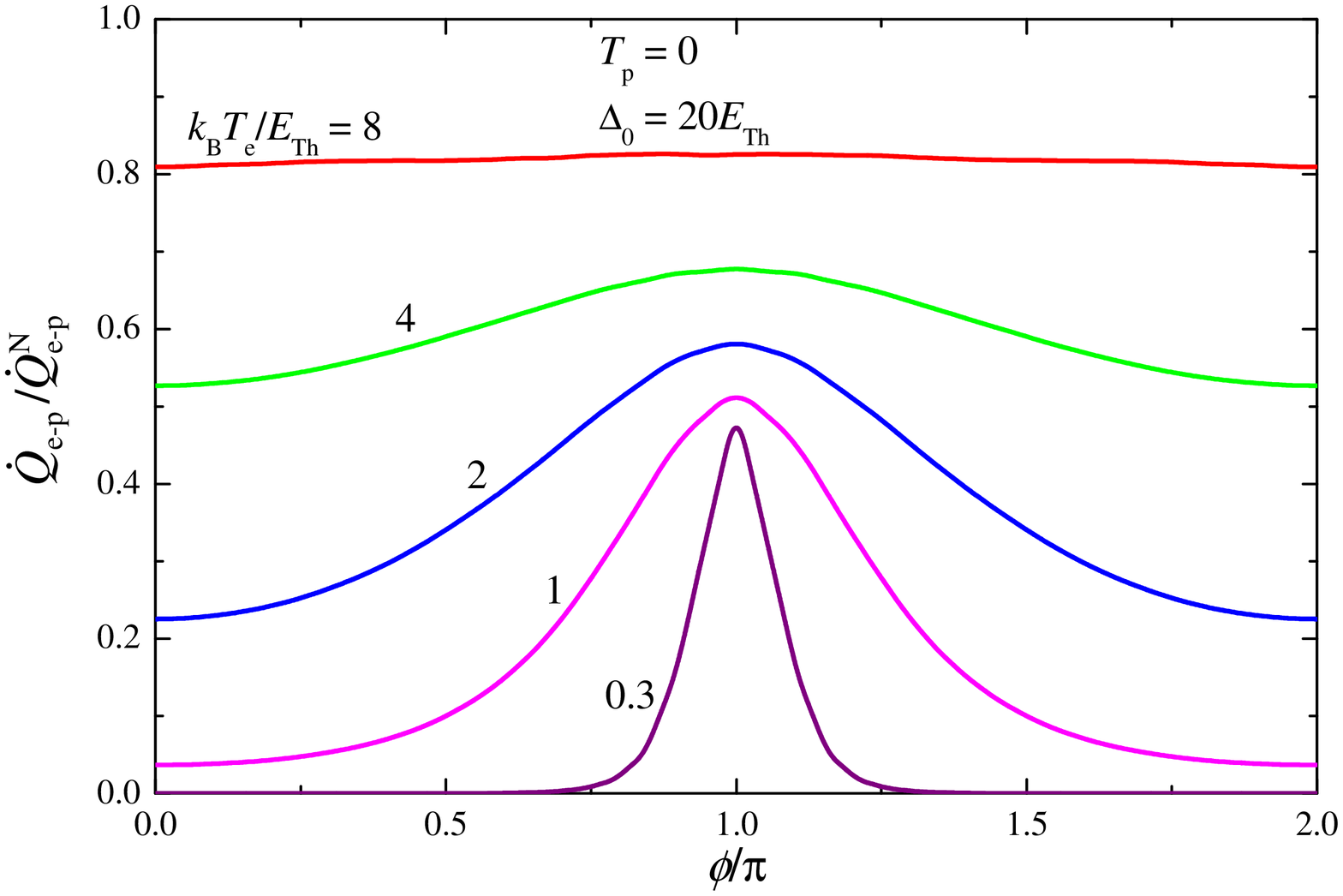}
\caption{(color online) Phase-dependent electron-phonon heat current
at a few different temperatures with $T_p=0$ and $\Delta=20 E_{\rm
Th}$. The closing minigap for $\phi \rightarrow \pi$ shows up as a
rapid increase of $\dot{Q}_{\rm e-p}$ at low temperatures. For $k_B
T_e$ somewhat larger than the minigap, the phase dependence becomes
much weaker although the overall heat current is still smaller than
in the normal state.} \label{fig:phasedep}
\end{figure}

Apart from the temperature and length of the junction, the nature of
the proximity effect is quite sensitive to the phase difference
$\phi$ between the superconductors. This sensitivity is mainly due
to the phase-dependent minigap in the density of states. This is
illustrated in Fig.~\ref{fig:phasedep}, where $\dot{Q}_{\rm e-p}$ is
plotted as a function of $\phi$. The strongest dependence takes
place naturally at temperatures close to the phase-dependent
minigap. However, some phase dependence is still remaining even at
higher temperatures.

In the long-junction regime $L \gg \xi_0$, $\phi=0$ and temperatures
up to some $10 E_T/k_B$ the temperature dependence of the power
$\dot{Q}_{\rm e-p}$ can be fitted rather well to a curve of the form
\begin{equation}\label{eq:tempdepfit}
\dot Q_{\rm e-p} = \dot Q_{\rm e-p}^N \exp(-T^*/T),
\end{equation}
where $T^*=c E_{\rm Th}/k_B$ is a temperature scale describing the
proximity-induced variations. For the prefactor we get $c\approx
3.5\dots 3.7$, depending on how large temperatures are included in
the fit. This behavior resembles that of bulk
superconductors,\cite{timofeev09} but where the gap $\Delta$ is
replaced by $T^*$. However, in the SNS case the exponential behavior
extends to comparatively larger temperatures than in bulk
superconductors where one has to assume $k_B T \ll \Delta$ for the
law to be valid.

As seen in Fig.~\ref{fig:phasedep}, $\dot Q_{\rm e-ph}$ becomes
independent of the phase at temperatures larger than the zero-phase
minigap. In this case the temperature dependence can again be
expressed with Eq.~\eqref{eq:tempdepfit}. However, for low
temperatures the effective temperature scale becomes of the order of
the phase-dependent minigap $E_g(\phi)$.

\section{Experimental determination}
\begin{figure}[t]
\centering
\includegraphics[width=\columnwidth,clip]{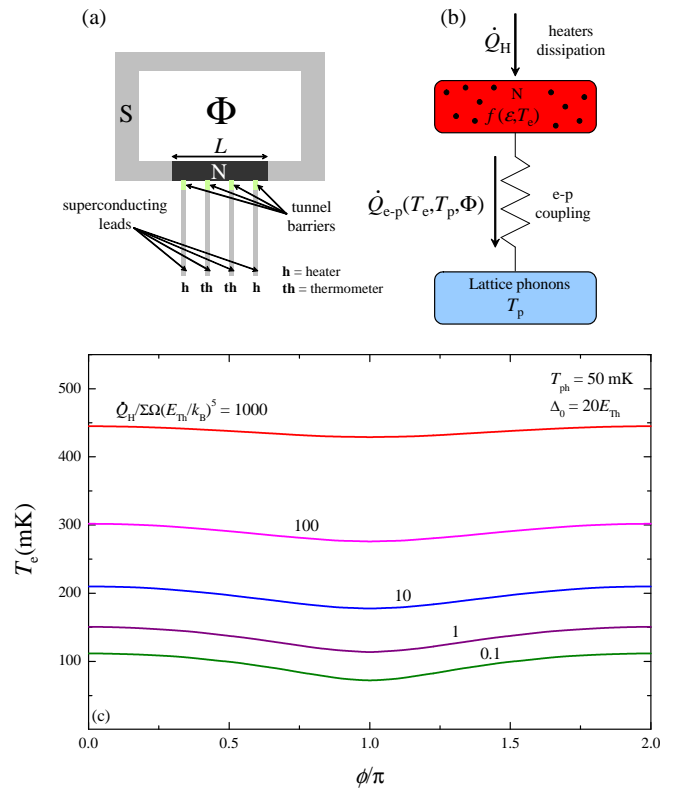}
\caption{(color online) (a) Scheme of a possible experimental setup.
A radio-frequency superconducting quantum interference device
(rf-SQUID) containing a SNS junction is threaded by a magnetic flux
$\Phi$ which allows to tune the electron-phonon interaction in the N
region. Additional superconducting electrodes tunnel-coupled to the
N wire allow to probe the averaged phase-dependent $\dot{Q}_{\rm
e-p}(T_e,T_p,\phi)$. These serve both as heaters ({\bf h}) and
thermometers ({\bf th}). (b) Sketch of the thermal model of the N
wire (see text). (c) Steady state electron temperature inside the
SNS junction as a function of the phase for a few driving powers at
a phonon temperature $T_{p}=E_{\rm Th}/(2k_B)$. For this curve, we
chose $E_{\rm Th}/k_B=100$ mK (see text).} \label{fig:heatbalance}
\end{figure}

A possible experimental setup to measure the effects discussed above
is shown in Fig.~\ref{fig:heatbalance}(a). The structure can be
fabricated through standard lithography techniques.\cite{giazotto06} %through standard electron beam lithography combined with
%shadow-mask evaporation and oxidation of thin-film metals.
The device consists of a radio frequency superconducting quantum
interference device (SQUID),\cite{baselmans2001,angers} where a
superconducting loop is interrupted by a N wire of length $L$. The
strength of the electron-phonon interaction in N can be modulated
periodically by an externally applied magnetic field, which gives
rise to a total flux $\Phi$ through the loop area. Neglecting the
inductance of the superconducting loop, the phase difference becomes
then $\phi=\Phi/(h/2e)$. The SQUID allows magneto-electric
characterization of the SNS junction, and thus also the
determination of some of the relevant parameters of the N wire. For
instance, the Thouless energy of the junction can be extracted from
the temperature dependence of the SQUID critical supercurrent.
Alternatively, the phase can be controlled by an externally driven
current in the case when the superconductors are not connected.
However, in this cases only phases $\phi \approx -\pi/2 \dots \pi/2$
can be accessed. The N region is connected to four additional
superconducting electrodes through oxide barriers, so to realize
normal metal-insulator-superconductor (NIS) tunnel junctions. The
NIS junctions are used to heat (or eventually to cool) the electrons
in N, and as sensitive thermometers to measure
$T_{\text{e}}$.\cite{giazotto06} Both the NS contacts and the NIS
junctions provide nearly ideal thermal isolation of the N region and
therefore in the following we neglect the thermal conductance
through the superconductors. One further advantage of using tunnel
junctions directly connected to N stems from the fact that the state
of the electron system and the superconducting correlations induced
by proximity from S electrodes will be virtually unaffected by the
presence of tunnel-coupled probes.\cite{giazotto04} The basic
requirements for this are that the currents driven through the NIS
junctions are much smaller than the SNS critical current, and that
the tunnel junction resistances are much larger than the resistance
of the normal metal. These requirements are easy to satisfy in
practice.

Figure \ref{fig:heatbalance}(b) shows a sketch of the relevant
thermal model of the wire in the proposed setup.\cite{timofeev09}
Upon heating electrons with a constant power $\dot{Q}_H$ provided by
the NIS heaters the steady-state $T_{\text{e}}$ established in N
will depend on the energy relaxation mechanisms occurring in the
wire. In metals at low lattice temperature (typically below 1 K),
the main relaxation mechanism is related to electron-phonon
interaction which in the present setup is strongly phase dependent.
In practice, what will be measured is the power flowing between the
electron and phonon systems averaged over the wire volume $\Omega$,
i.e., $\dot{Q}_{\text{\rm e-p}}(T_{\text{e}},T_{\text{p}},\phi)$. At
fixed lattice temperature the predicted steady-state phase-dependent
$T_{\text{e}}$ thus follows from the solution of the thermal-balance
equation
\begin{equation}
\dot{Q}_H+\dot{Q}_{\text{\rm
e-p}}(T_{\text{e}},T_{\text{p}},\phi)=0.
\end{equation}
The resulting electron temperature $T_e$ can then be probed with the
NIS thermometers.

The result of a such a heat balance is plotted in
Fig.~\ref{fig:heatbalance} which displays the steady-state electron
temperature $T_e$ for some values of $\dot{Q}_H$ as the phase is
varied and the bath temperature is $T_p=50$ mK. To illustrate the
strength of the effect, we chose a typical $E_{\rm Th}/k_B=100$ mK.
For each $\dot{Q}_H$ the electronic temperature is strongly
modulated by the phase, and it decreases by an increasing $\phi$ due
to the enhancement of the electron-phonon interaction for phase
difference close to $\pi$ (see Fig.~5). In particular, for
$\dot{Q}_H=0.1\Sigma\Omega(E_{\rm Th}/k_B)^5$, variation of $T_e$ of
the order of 40 mK (0.4 $E_{\rm Th}$) can be achieved by varying the
phase (flux), while even for the large injected power
$\dot{Q}_H=100\Sigma\Omega(E_{\rm Th}/k_B)^5$ the variation
amplitude is somewhat below 30 mK (0.3 $E_{\rm Th}$).

Let us discuss some of the typical materials parameters for the
proposed measurement. By choosing, for instance, aluminum (Al) as S
material (with $\Delta_0= 200\,\mu$eV) and copper (Cu) as N wire
(with $D=0.01$ m$^2$s$^{-1}$ and $\Sigma=2\times 10^9$
Wm$^{-3}$K$^{-5}$, see Ref.~\onlinecite{giazotto06}) and length
$L=0.8\,\mu$m, we get $\Delta_0/E_{\text{T}}= 19$ and $E_{\rm
Th}/k_B = 10$ $\mu$eV $=120$ mK. Moreover, if we assume that the
wire is 50 nm thick and 400 nm wide, the "unit power" in
Fig.~\ref{fig:heatbalance} is $\Sigma \Omega (E_{\rm Th}/k_B)^5
=0.8$ fW.

%The phase dependence of the electron-phonon energy current can be
%used to directly probe the proximity-induced changes in the
%electron-phonon coupling. Such probing can be obtained for example
%by heating the junction with a constant power $\dot{Q}_H$, and
%measuring the resulting electron temperature. In this case the final
%electron temperature will be obtained from the heat balance equation
%\begin{equation}
%\dot{Q}_H=\dot{Q}_{\rm e-p}(T_e,T_p,\phi).
%\end{equation}
%The result of such a heat balance is plotted in
%Fig.~\ref{fig:heatbalance}, along with the proposed measurement
%setup.

\section{Conclusions}

In bulk superconductors, the presence of the energy gap leads to a
strong suppression of the heat transport between the electron and
phonon systems.\cite{timofeev09} In this paper we show that a
similar effect can be expected for normal metals in close proximity
to the superconductors and suggest how it can be probed.
Additionally, the heat current between the two systems can be
controlled in situ by changing the phase difference between two
superconducting contacts. The suppression of the heat conductance is
especially relevant in thermal radiation detectors --- for example
in the Proximity Josephson Sensor (PJS) suggested in
Ref.~\onlinecite{giazotto08}. This is because of two factors: First,
the weaker the thermal relaxation, the stronger is the temperature
rise for a given power of radiation. Secondly, the main cause of
noise in these devices is related to the strongest thermal
relaxation channel, which typically is the electron-phonon coupling.
Reducing this coupling will therefore also reduce the noise. Both of
these factors will improve the sensitivity of such thermal
detectors.

\section*{Acknowledgements}

We thank Pauli Virtanen for discussions and help with the numerics.
TTH was supported by the Academy of Finland and FG partially by the
NanoSciERA "NanoFridge" project of the EU. TTH acknowledges the
hospitality of the Kavli Institute of Nanotechnology at the Delft
University of Technology, where part of this work was carried out.

\section{Appendix}
The self-energy for the electron-phonon scattering in the case of
thermal phonons at temperature $T_{p}$ and with the general Keldysh
Green's function for the electrons is given by\cite{kopnin}
\begin{equation*}
\check{\Sigma}_{\rm e-p} = \begin{pmatrix}\hat{\Sigma}^R &
\hat{\Sigma}^K \\ 0& \hat{\Sigma}^A \end{pmatrix},
\end{equation*}
where
%\begin{widetext}
\begin{align*}
\begin{split}
\hat{\Sigma}^{K}=\lambda \int \frac{d\omega}{4} \omega^2 {\rm
sgn}(\omega) \bigg[&\coth\left(\frac{\omega}{2 k_B T_p}\right)
\hat{g}^K(\epsilon+\omega)\\&-\hat{A}(\epsilon+\omega)\bigg]
\end{split}\\
\begin{split}
\hat{\Sigma}^{R/A}=\lambda \int \frac{d\omega}{4} \omega^2
\sgn(\omega) \bigg[&\coth\left(\frac{\omega}{2 k_B T_p}\right)
\hat{g}^{R/A}(\epsilon+\omega)\\ &\mp \frac{1}{2}
\hat{g}^K(\epsilon+\omega)\bigg].
\end{split}
\end{align*}
%\end{widetext}
Here $\hat{A}=\hat{g}^R-\hat{g}^A$ is the spectral function and
$\hat{g}^{R,A,K}$ are the retarded, advanced and Keldysh Green's
functions, respectively. From these self-energies, the collision
integral appearing in Usadel equation is
\begin{equation*}
\check{I}_{\rm e-p} =
\begin{pmatrix} \hat{I}^R & \hat{I}^K \\ 0 & \hat{I}^A \end{pmatrix}= \left[\check{g},\check{\Sigma}_{\rm e-p}\right].
\end{equation*}
We are mostly interested in the Keldysh part, which has two
components: $I_L={\rm Tr}\left[\hat{I}^K\right]/2$ and $I_T={\rm
Tr}\left[\hat{\tau}_3 \hat{I}^K\right]/2$. In the following we
parameterize the Green's function by $\hat{g}^K=\hat{g}^R
\hat{h}-\hat{h} \hat{g}^A$, $\hat{h}=f_L \hat{\tau}_0 + f_T
\hat{\tau}_3$, and
\begin{equation*}
\hat{g}^R = \begin{pmatrix} g & f e^{i \phi} \\ -f e^{-i\phi} &
-g\end{pmatrix}.
\end{equation*}
Moreover, the normalization condition implies $g^2-f^2=1$. The
Advanced Green's function can then be obtained by
$\hat{g}^A=-\hat{\tau}_3 (\hat{g}^R)^\dagger \hat{\tau}_3$. Note
that inside a proximity structure, all the three variables, $g$, $f$
and $\phi$ are in general complex-valued. Moreover, we define
$N={\rm Re}[g]$ (local density of states) and the real quantities
\begin{equation}
\begin{split}
F_c &= {\rm Re}[f\cos(\phi)],\quad
F_s = {\rm Re}[f\sin(\phi)]\\
\tilde{F}_c &= {\rm Im}[f\cos(\phi)],\quad \tilde{F}_s = {\rm
Im}[f\sin(\phi)].
\end{split}
\label{eq:ffuncs}
\end{equation}
After some lengthy but straightforward algebra, the collision
integrals become
\begin{widetext}
\begin{equation}
\begin{split}
I_T =& \lambda \int d\omega \omega^2 \sgn(\omega)
\bigg\{[F_c(\epsilon+\omega)\tilde{F}_s(\epsilon)-F_s(\epsilon+\omega)\tilde{F}_c(\epsilon)]\left[\coth\left(\frac{\omega}{2
k_B T_{p}}\right) f_L(\epsilon+\omega)-1\right]\\
&+[F_s(\epsilon)
\tilde{F}_c(\epsilon+\omega)-F_c(\epsilon)\tilde{F_s}(\epsilon+\omega)]\coth\left(\frac{\omega}{2
k_B T_{p}}\right) f_L(\epsilon)\\& +
[N(\epsilon)N(\epsilon+\omega)+\tilde{F}_c(\epsilon+\omega)
\tilde{F}_c(\epsilon)+\tilde{F}_s(\epsilon+\omega)
\tilde{F}_s(\epsilon)] \coth\left(\frac{\omega}{2 k_B T_{p}}\right)
[f_T(\epsilon+\omega)-f_T(\epsilon)]\\&
+N(\epsilon)N(\epsilon+\omega)[f_L(\epsilon)f_T(\epsilon+\omega)+f_L(\epsilon+\omega)f_T(\epsilon)]\bigg\}
\end{split}
\end{equation}
and
\begin{equation}
\begin{split}
I_L=& \lambda \int d\omega \omega^2 \sgn(\omega) \bigg\{[N(\epsilon)
N(\epsilon+\omega)-F_c(\epsilon)F_c(\epsilon+\omega)-F_s(\epsilon)F_s(\epsilon+\omega)]\times\\&\times\left[\coth\left(\frac{\omega}{2
k_B T_{p}}\right)
[f_L(\epsilon+\omega)-f_L(\epsilon)]-1+f_L(\epsilon)f_L(\epsilon+\omega)\right]\\&+
[F_s(\epsilon)\tilde{F}_c(\epsilon+\omega)-F_c(\epsilon)\tilde{F}_s(\epsilon+\omega)]\left[f_L(\epsilon)+\coth\left(\frac{\omega}{2
k_B T_{p}}\right)\right]f_T(\epsilon+\omega)\\&+
[F_s(\epsilon+\omega)\tilde{F}_c(\epsilon)-F_c(\epsilon+\omega)\tilde{F}_s(\epsilon)]\left[f_L(\epsilon+\omega)-\coth\left(\frac{\omega}{2
k_B T_{p}}\right)\right]f_T(\epsilon)
\\&+[N(\epsilon)N(\epsilon+\omega)-\tilde{F}_c(\epsilon)\tilde{F}_c(\epsilon+\omega)-\tilde{F}_s(\epsilon)\tilde{F}_s(\epsilon+\omega)]
f_T(\epsilon)f_T(\epsilon+\omega)\bigg\}.
\end{split}
\end{equation}
These simplify considerably in the quasiequilibrium limit where
$f_T=0$ and $f_L=\tanh[\epsilon/(2k_B T)]$. In that case one obtains
the collision integral presented in Eq.~\eqref{eq:quasieqcollil}.
\end{widetext}


\begin{thebibliography}{99}
\bibitem{numnote} In order to find the kernel $K(\epsilon,\omega)$,
we parameterize Eq.~\eqref{eq:usadel} with the Riccatti-type
parametrization detailed for example in Ref.~\onlinecite{hammer07}.
This parameterization, contrary to the often employed
$\theta$-parameterization (see, e.g., Ref.~\onlinecite{virtanen07}),
can access all values of the phase difference $\phi$ between the
superconductors.
\end{thebibliography}
\end{document}